\documentclass[a4paper,twocolumn]{esapub2005} % European paper
\usepackage{times,graphicx}
\usepackage{natbib}

\title{The pulsations and potential for seismology of B stars}

\author[1,2]{C. Aerts} \affil[1]{Instituut voor Sterrenkunde, Katholieke
Universiteit Leuven, Celestijnenlaan 200 D, B-3001 Leuven,
Belgium,conny@ster.kuleuven.be} \affil[2]{Department of Astrophysics, University
of Nijmegen, P.O. Box 9010, 6500 GL Nijmegen, The Netherlands}

\begin{document}


\maketitle

\begin{abstract}
We review the nature of the oscillations of main-sequence and supergiant stars
of spectral type B. Seismic tuning of the interior structure parameters of the
$\beta\,$Cep stars has been achieved since three years. The results are based on
frequencies derived from long-term monitoring and progress in this area is
rapid.  Oscillations in mid-B stars as well as Be stars are well established by
now, but we lack good mode identification to achieve seismic modelling. We
provide recent evidence of g-mode pulsations in supergiant B stars. The
spherical wavenumbers of their modes are yet unidentified, preventing seismic
probing of such evolved hot stars at present. Improving the situation for the
three groups of g-mode oscillators requires multi-site long-term high-resolution
spectroscopy in combination with either space photometry or ground-based
multicolour photometry. The CoRoT programme and its ground-based programme will
deliver such data in the very near future.
\end{abstract}

\section{Introduction}

A large fraction of the stars of spectral type B is known to be variable.
Since more than a century now, these variables have been divided in different
classes, according to their periods and morphology of the lightcurves. In this
review, we concentrate on those classes of variable B stars with established
periodic variability resulting from stellar oscillations and situated near or
above the main sequence. This concerns the classes of the $\beta\,$Cep stars,
the slowly pulsating B stars, the pulsating Be stars and the pulsating
supergiant B stars. For a review on the oscillations of subdwarf B stars, we
refer to the paper by Fontaine (these proceedings).

Large inventories of pulsating B stars were established during the first part of
the 20{\it th\/} century. These were mainly based on photographic spectroscopy
(see [1] for one of the earliest review papers).  The
introduction of photo-electric photometry in the second half of the 20{\it th\/}
century allowed much larger systematic survey campaigns, resulting in fainter
class members among them cluster stars. The Hipparcos mission subsequently
allowed the discovery of more than 100 bright periodic B stars [2].
Still today, new pulsating B stars are found, mainly from large-scale
surveys, as we will discuss below for each class separately. These early survey
works resulted in a fairly good statistics of the frequencies and amplitudes of
the oscillations, but not beyond that.

As of the 1970s, the research of pulsating B stars extended towards the area of
mode identification from observations. The motivation for this was that, at that
time, the samples of pulsating B stars were large enough to delineate the
observational instability strips, but no instability mechanism was known to
explain the oscillations. Identification of the mode wavenumbers $(\ell,m)$
could therefore help to discover such a mechanism and to understand the
mode selection.  Mode identification was first mainly attempted from multicolour
photometry using the method introduced by [3] and based on previous
theoretical works by [4] and [5], [6]. The
degree of the oscillation modes can be identified from amplitude ratios and/or
phase differences (see, e.g., [7] for a review of this method and
[8] for a recent improvement).  Later on, from the mid 1980s,
the possibility of performing high-resolution spectroscopy emerged from improved
instrumental technology. This, in combination with the suggestion of [9]
that one can compute theoretical line profiles for various kinds of
nonradial oscillations, initiated a series of still ongoing efforts to obtain
high spatial- and time-resolution spectroscopic observations of pulsating stars
B with the specific aim to perform mode identification. 

Meanwhile, the instability mechanism is well known. It is the $\kappa$-mechanism
acting in the partial ionisation zones of the iron-like elements (see [10] for
an excellent review). The mode selection, however, is still totally unknown to
us.

It was only a few years ago that accurate enough frequencies, combined with
unambiguous mode identification, became available for several nonradial modes
in a few selected B stars which had been monitoring since many years. In this
paper, we report on the current status of B star asteroseismology, highlighting
the recent successes in the seismic interpretation of the interior structure
parameters of the $\beta\,$Cep stars, and pointing out the difficulties yet to
overcome to achieve the same success for other B-type pulsators.

\section{$\beta\,$Cep stars}
\begin{figure*}
\begin{center}
\rotatebox{0}{\resizebox{5.5cm}{!}{\includegraphics{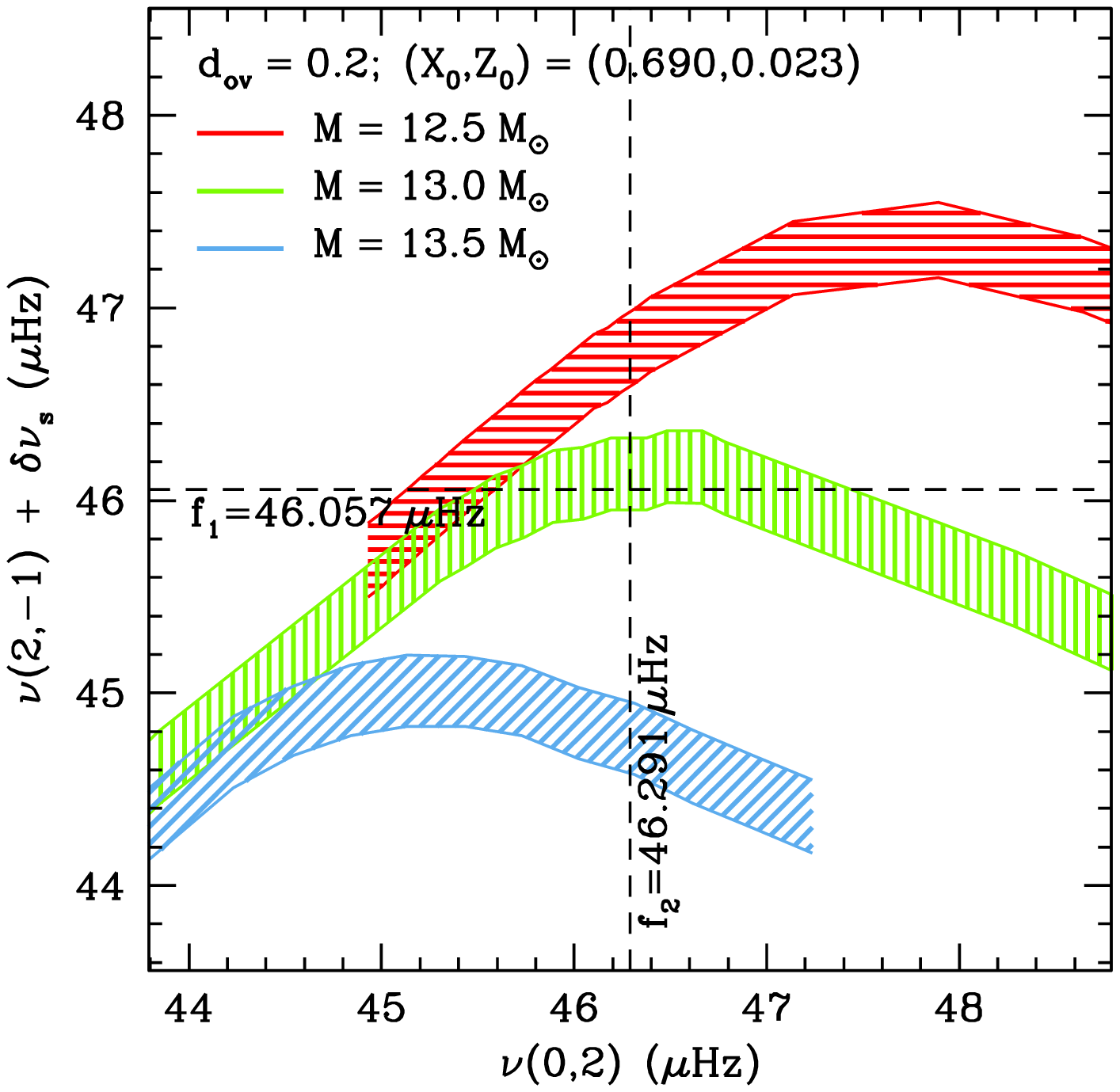}}} 
\rotatebox{0}{\resizebox{5.5cm}{!}{\includegraphics{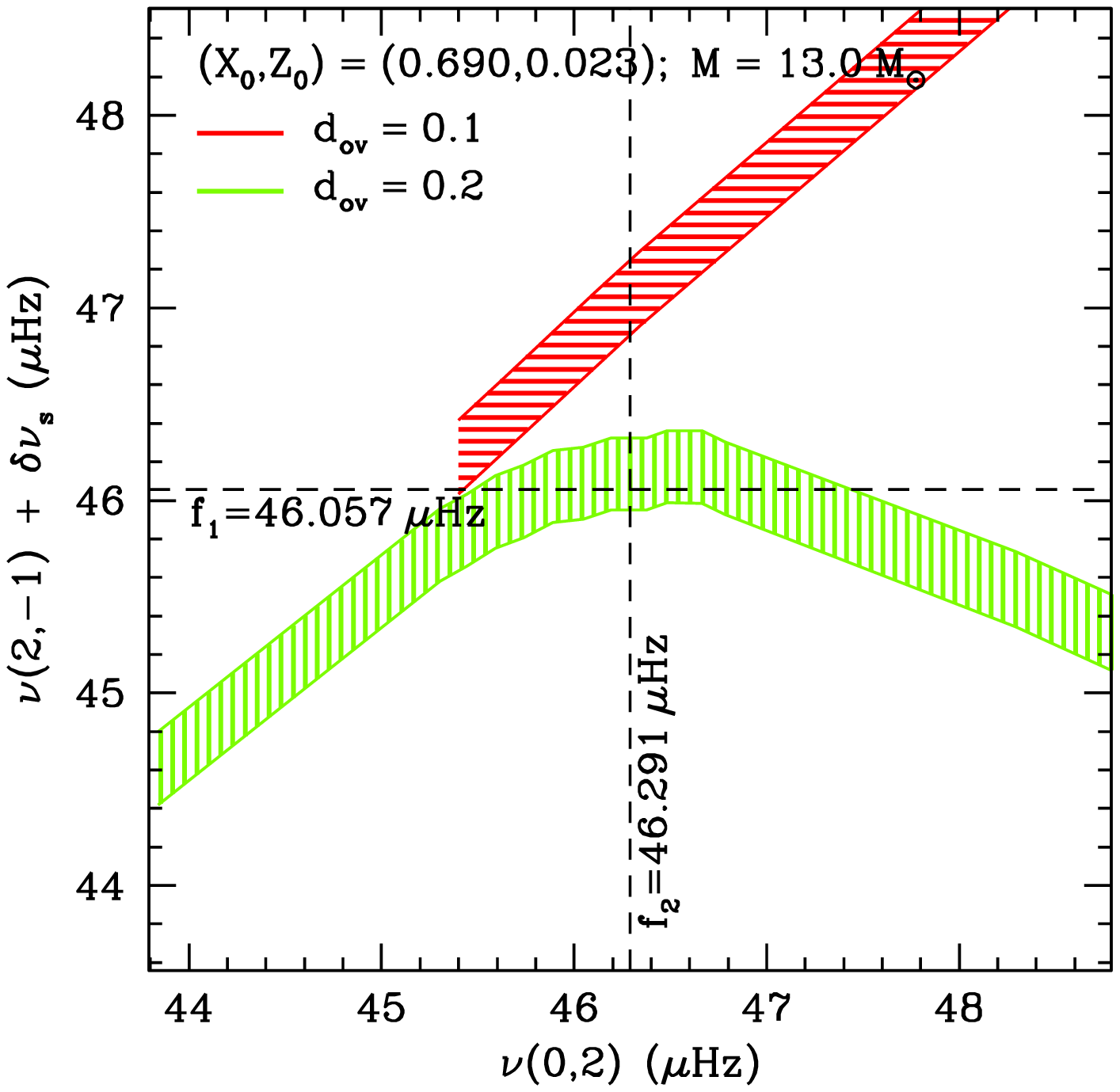}}} 
\rotatebox{0}{\resizebox{5.5cm}{!}{\includegraphics{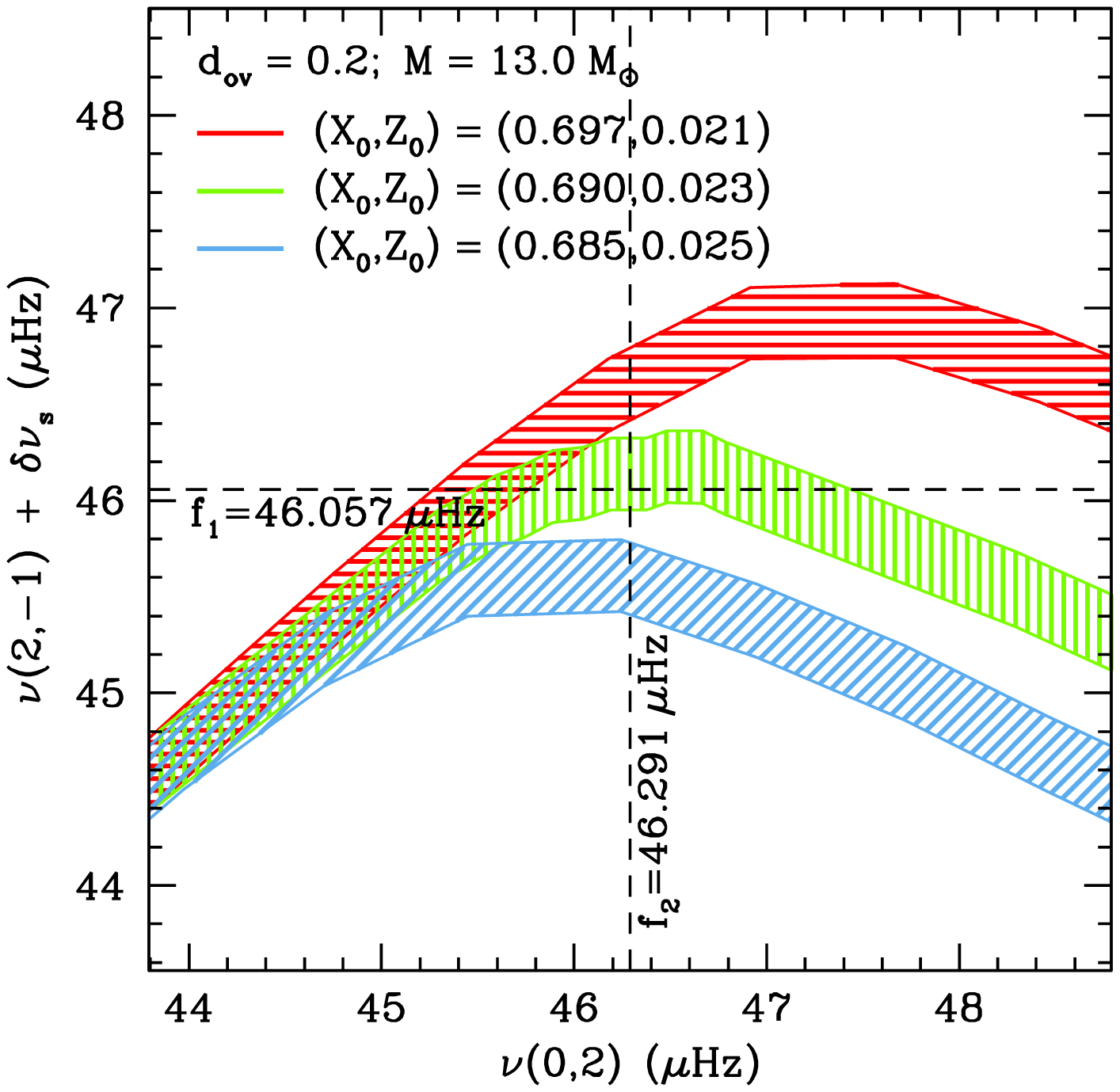}}} 
\end{center}
\caption[] {The variation of theoretical frequencies, computed for models
appropriate for the $\beta\,$Cep star $\beta\,$CMa, with the stellar parameters
$M$ (left), $d_{\rm ov}$ (middle) and $Z$ (right).  In each panel, two of the
parameters are fixed to visualise the effect of the other remaining parameter on
the frequencies. The first radial overtone frequencies are plotted as abcissae
and the $\ell=2$ g$_1$ mode frequency, corrected for the rotational splitting,
are plotted as ordinates. The dashed lines indicate the two observed
frequencies, i.e.\ the point where they interset indicates a perfect match of
theoretical and observed frequencies. Each area represents an evolutionary
track whose width is due to the uncertainty in the measured rotational
splitting.  Figure reproduced from [32] with permission from
A\&A and from the authors.}
\label{anwesh}
\end{figure*}

The $\beta\,$Cep stars are a well-established group of near-main sequence
pulsating stars.  They have masses between 8 and about 18\,M$_\odot$ and
oscillate in low-order p and g modes with periods between about 2 and 8\,h
excited by the $\kappa\,$mechanism acting in the partial ionisation zones of
iron-group elements [11]. The agreement between observed $\beta\,$Cep stars and
the theoretical instability strip is very satisfactory for the class as a whole,
although the blue part of the strip is not well populated [12]. Most of the
$\beta\,$Cep stars show multiperiodic light and line profile variations.  The
majority of the $\beta\,$Cep stars rotate at only a small fraction of their
critical velocity. An recent overview of the observational properties of the
class is available in [13].

Recently, numerous new candidate members have been found from large-scale
surveys, in the LMC and SMC [14] as well as in our own Galaxy [15],
[16]. Assuming that all these faint variable stars are indeed $\beta\,$Cep stars
more than doubles the number of class members to over 200. The occurrence of so
many $\beta\,$Cep stars in environments with very low metallicity implies new
unanticipated challenges to the details of the mode excitation, which relies
heavily on the iron opacity.

The amplitudes and frequencies of the $\beta\,$Cep stars seem quite stable,
although very few dedicated long-term studies are available.  The B2III star
12\,Lac, e.g., was known to have six oscillation modes from photometry [17] and
these same modes were recovered in high-resolution spectroscopy more than a
decade later [18] and yet again, together with many more modes, in a recent
multisite campaign [19].  The B3V star HD\,129929, on the other hand, was
monitored during 21 years in 3-week campaigns from La Silla with one and the
same high-precision photometer attached to the 0.70-m Swiss telescope [20].
This also led to the detection of six independent oscillation modes,
with very small amplitude variability for the triplet frequencies only, if
any. Suggestions for evolutionary frequency changes from O-C diagrams have been
made, but we regard these as premature.

Significant progress in the detailed seismic modelling of the $\beta\,$Cep stars
has occurred since a few years. While such modelling was already attempted a
decade ago for the stars 16\,Lac [21] and 12\,Lac
[22], doubtful mode identification prevented
quantitative results. It took until the exploitation of the 21-yr single-site
multi-colour data set of the star HD\,129929 to discover that standard stellar
models are unable to explain that star's oscillation behaviour. Indeed, from the
modelling of three identified $m=0$ modes, [20] derived a core
overshoot parameter of $0.10\pm 0.05$\,H$_{\rm p}$ (with H$_{\rm p}$ the local
pressure scale height) and proved the star to undergo non-rigid internal
rotation from the splitting within an $\ell=2$ and an $\ell=1$ mode, with the
core rotating four times faster than the envelope. For details, we refer to
[23] and [24].

This modelling result was soon followed by the one derived for the B2III star
$\nu\,$Eri, which was the target of a 5-month multisite photometric and
spectroscopic campaign. Numerous new frequencies were found and identified
compared to the four known before the start of the campaign [25], [26], [27].
The modelling was done by two independent teams using different evolution and
oscillation codes. This led to different results depending on the number of
fitted $m=0$ components (three $m=0$ modes were fitted by [28] while four by
[29]).  The main and far most important conclusion was, however, the same for
both studies: current seismic models do not predict all the observed modes of
$\nu\,$Eri to be excited. One needs a factor four enhancement in the iron
opacity, either locally in the driving region, or globally in the star, to solve
this excitation problem. This led to the suggestion to include radiative
diffusion in the models to solve this outstanding issue, in analogy to the
subdwarf B pulsators [30]. Promising first attempts to compute main-sequence
B-star models including diffusion were made by [31]. They found that the
diffusion effects do not alter the frequency values in a significant way, but
have indeed the potential to solve $\nu\,$Eri's excitation problem (or, better
phrased: our inability to explain its mode excitation \ldots).

Meanwhile, two more $\beta\,$Cep stars were modelled seismically, each of them
having two well-identified frequencies. The example of $\beta\,$CMa 
is illustrative of the power of asteroseismology: having two
well-identified oscillation modes in a slow rotator is sufficient to derive a
quantitative estimate of the core overshoot parameter, which was found to be
$d_{\rm ov}= 0.15\pm 0.05$\,H$_{\rm p}$ for this somewhat evolved B2III
$\beta\,$Cep star. The way this is achieved, is illustrated nicely in
Fig.\,\ref{anwesh}, taken from the paper by [32]. Because the
frequency spectra of $\beta\,$Cep stars are so sparse for low-order p and g
modes, one does not have many degrees of freedom to fit the well-identified
modes. This is why we can put limits on internal structure parameters as shown
in Fig.\,\ref{anwesh}, of course assuming that the input physics of the models
is the correct one. A similar, but less stringent constraint was derived for the
B2IV star $\delta\,$Ceti from a combination of MOST space photometry and
archival ground-based spectroscopy [33].

Additional multisite campaigns have been done for the stars $\theta\,$Oph [34],
[35], 12\,Lac [19] and V\,2052\,Oph (Handler, unpublished). These have a
somewhat higher projected rotation velocity, and it would be interesting to know
if the range of values found so far for the core overshoot parameter and the
level of non-rigidity of the internal rotation remains valid for them. The
modelling is ongoing at present.

\section{Slowly pulsating B stars}

The term ``slowly pulsating B stars'' (SPB stars) was introduced by [36], after
years of photometric monitoring of variable mid-B stars with multiperiodic
brightness and colour variations.  After a few years, the Hipparcos mission led
to a tenfold increase in the number of class members [2]. 
Subsequent huge long-term multicolour photometric and high-resolution
spectroscopic follow-up campaigns concentrated on the brightest new class
members found from Hipparcos [37], [38] and
resulted in a much better understanding of the pulsational and rotational
behaviour of the class members [39].  Accurate frequencies and mode
identification are available for some 15 members [40], [41]. 
The mode identification results are in excellent agreement with
theoretical computations made by [42] predicting mainly dipole modes
to be excited.  All confirmed SPB stars are slow rotators [39].

\begin{figure*}
\begin{center}
\rotatebox{0}{\resizebox{14.cm}{!}{\includegraphics{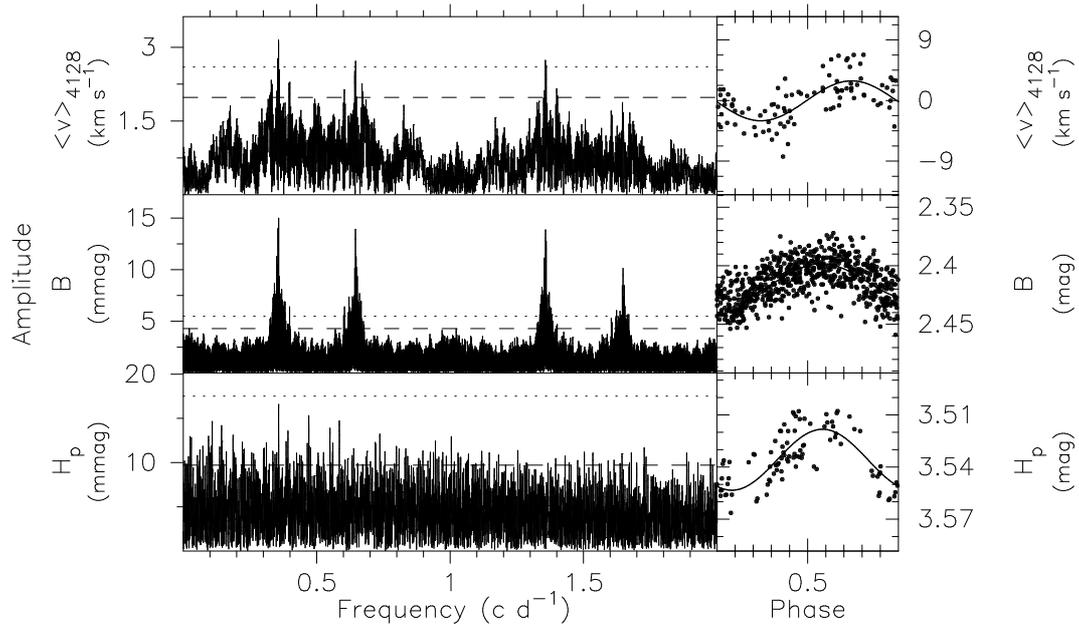}}} \end{center}
\caption[] {The frequency spectra of Geneva $B$, Hipparcos, and radial velocity
data derived from the Si\,\textsc{ii} 4128\,\AA\ line of the SPB star
HD\,74195. The horizontal dashed line indicates the 1\% false-alarm probability
and the dotted one the 3.7 S/N ratio level. Figure reproduced from [40]
with permission from A\&A and from the authors.}
\label{ovel}
\end{figure*}

\begin{figure*}
\begin{center}
\rotatebox{270}{\resizebox{11cm}{!}{\includegraphics{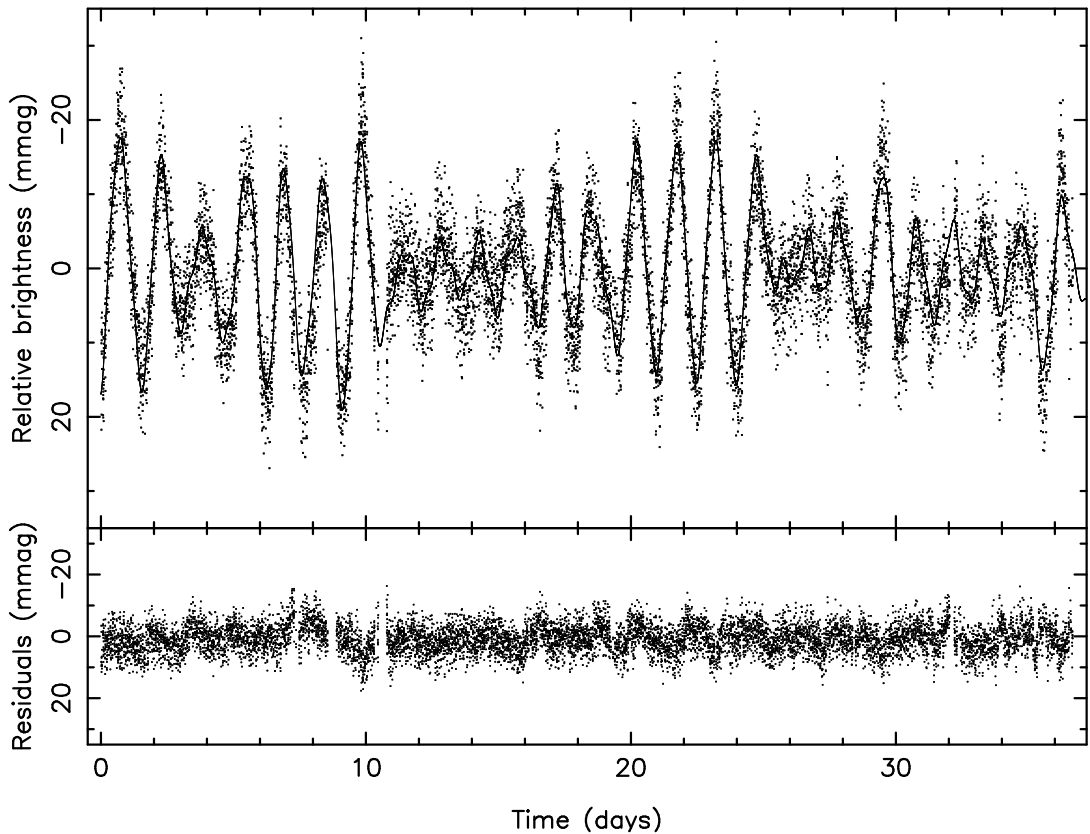}}} 
\end{center}
\caption[] {The MOST light curve of the SPB HD\,163830 (upper panel, dots) and
the best fit based on the 21 significant frequencies (upper panel, full
line). The residuals after subtraction of the fit are shown in the lower
panel. Figure reproduced from [43] with permission from the ApJ
and from the authors.}
\label{hd163830}
\end{figure*}

In Fig.\,\ref{ovel} we show as an illustration the frequency spectrum of the
Geneva $B$ and Hipparcos light, and radial velocity variations of the brightest
among the SPB stars, $o\,$Vel (B3IV).  Despite the long-term monitoring of
almost two decades in photometry, [40] found only four independent frequencies
for this star. This is typical for single-site ground-based data of
main-sequence stars with gravity modes, because the latter have periodicities
ranging from 0.8 to 3\,d. This leads to severe alias problems, as illustrated in
Fig.\,\ref{ovel} where the confusion between frequencies $f$ and $1-f$ is
prominent.  Only with multisite data, or, even better, with uninterrupted data
from space, can one avoid such confusion. This is illustrated nicely by the MOST
light curve (reproduced in Fig.\,\ref{hd163830}) of the new SPB star HD\,163830
discovered by that mission [43]. This lightcurve implied a five-fold of the
number of gravity modes in one star compared to the best ground-based datasets
for such pulsators.

As for the $\beta\,$Cep stars, numerous new SPB stars (some 70) were discovered
in the Magellanic Clouds from OGLE and MACHO data [14].  The number of class
members is therefore about 200 at the time of writing (assuming all the
Magellanic Clouds variables to have been classified correctly).  Trustworthy
mode identification is only available for the highest-amplitude frequency of a
handful of SPB stars, however, and it concerns only the spherical wavenumbers of
the dominant mode [41].  This is why seismic tuning of the interior structure of
SPB stars has not been achieved so far.

\section{Pulsating Be stars}
\begin{figure*}
\begin{center} 
%\parbox{12.5cm}{ \parbox{3.8cm}{
%\psfig{figure=rivi1.ps,width=4.1cm,angle=0,clip=t}
%\psfig{figure=rivi2.ps,width=4.1cm,angle=0,clip=t} }
%\parbox{4.1cm}{
%\psfig{figure=rivi3.ps,width=4.1cm,angle=0,clip=t}
%\psfig{figure=rivi4.ps,width=4.1cm,angle=0,clip=t} }
%\parbox{4.1cm}{
%\psfig{figure=rivi5.ps,width=4.1cm,angle=0,clip=t}
%\psfig{figure=rivi6.ps,width=4.1cm,angle=0,clip=t} } }
\end{center}
\caption[]{Line profile variations in Be stars, with increasing $v\sin i$ for
FW\,CMa (top left, $v\sin i=40\,$km\,s$^{-1}$), $\omega\,$CMa (bottom left,
$v\sin i=100\,$km\,s$^{-1}$), $\mu\,$Cen (top middle, $v\sin
i=155\,$km\,s$^{-1}$), DX Eri (bottom middle, $v\sin i=180\,$km\,s$^{-1}$),
$\alpha\,$Eri (top right, $v\sin i=225\,$km\,s$^{-1}$), $\eta\,$Cen (bottom
right, $v\sin i=350\,$km\,s$^{-1}$). Data taken from [46].}
\label{rivi}
\end{figure*}

Be stars are Population~I B stars close to the main sequence that show, or have
shown in the past, Balmer line emission in their photospheric spectrum. This
excess is attributed to the presence of a circumstellar equatorial disk. See the
review on Be stars by [44] for general information on this rather inhomogeneous
class of stars.  Magnetic fields [45] and nonradial oscillations [46] have been
detected in some Be stars. It is unclear at present if these mechanisms are able
to explain a disk for the whole class of Be stars.

Be stars show variability on very different time scales and with a broad range
of amplitudes. [47] studied a subclass of the Be stars showing one dominant
period between 0.5 and 2\,d in their photometric variability, with amplitudes of
a few tens of a mmag which he termed the $\lambda\,$Eri variables.  He provided
extensive evidence of a clear correlation between the photometric period and the
rotational period of the $\lambda\,$Eri stars and interpreted that correlation
in terms of rotational modulation. When observed spectroscopically, several of
the $\lambda\,$Eri stars turn out to have complex line profile variations with
travelling sub-features similar to those observed in the rapidly rotating
$\beta\,$Cep stars, except for the much longer periods (days versus hours). This
rather seems to suggest oscillations as origin of this complex spectroscopic
variability.

Nonradial oscillations were already discovered in the Be star $\omega\,$CMa
[48], a star listed among the $\lambda\,$Eri variables in [47]'s list.  An
extensive summary of the detection of short-period line profile variations due
to oscillations in hot Be stars is provided in [46]. They monitored 27
early-type Be stars spectroscopically during six years and found 25 of them to
be line profile variables at some level. Some of their data are shown in a
grey-scale plot in Fig.\,\ref{rivi}. For several of their targets the
variability was interpreted in terms of nonradial oscillations with
$\ell=m=+2$. Almost all stars in the sample also show traces of outburst-like
variability rather than a steady star-to-disk mass transfer. The authors
interpreted the disk formation in terms of multimode beating in combination with
fast rotation.

The view on pulsating Be stars became more complicated when [49] introduced the
class of $\zeta\,$Oph variables. These are late-O type stars with clear complex
multiperiodic line profile variations which he attributed to high-degree
nonradial oscillations. They are named after the prototypical O9.5V star
$\zeta\,$Oph, whose rotation is very close to critical and whose photometric
variability was recently firmly established by the MOST space mission. [50]
disentangled a dozen significant oscillation frequencies in the 24-d photometric
light curve assembled from space. These frequencies range from 1 to 10\,d$^{-1}$
and clearly indicate the star's relationship to the $\beta\,$Cep stars.

Multiperiodic oscillations were recently also reported in the rapidly rotating
B5Ve star HD 163868 from a 37-d MOST light curve.
[51] derived a rich frequency
spectrum, with more than 60 significant peaks, resembling that of an SPB star
and termed the star an SPBe star in view of its Be nature. They interpreted the
oscillation periods between 7 and 14\,h as high-order prograde sectorial g~modes
and those of several days as Rossby modes (e.g. [52] for a good
description of such modes). There is remaining periodicity above 10\,d which
cannot be explained at present. Finally, nonradial oscillations at low amplitude
were also detected in the bright B8Ve star $\beta\,$CMi [53].

As for the SPB stars, seismic modelling of the interior structure of Be stars
has not yet been achieved, in this case by lack of enough frequencies, 
of frequency accuracy, of unambiguous mode identification and of appropriate
stellar models for rapid rotators.

\section{Pulsating B supergiants}
\begin{figure*}
\begin{center}
\rotatebox{90}{\resizebox{9.cm}{15cm}{\includegraphics{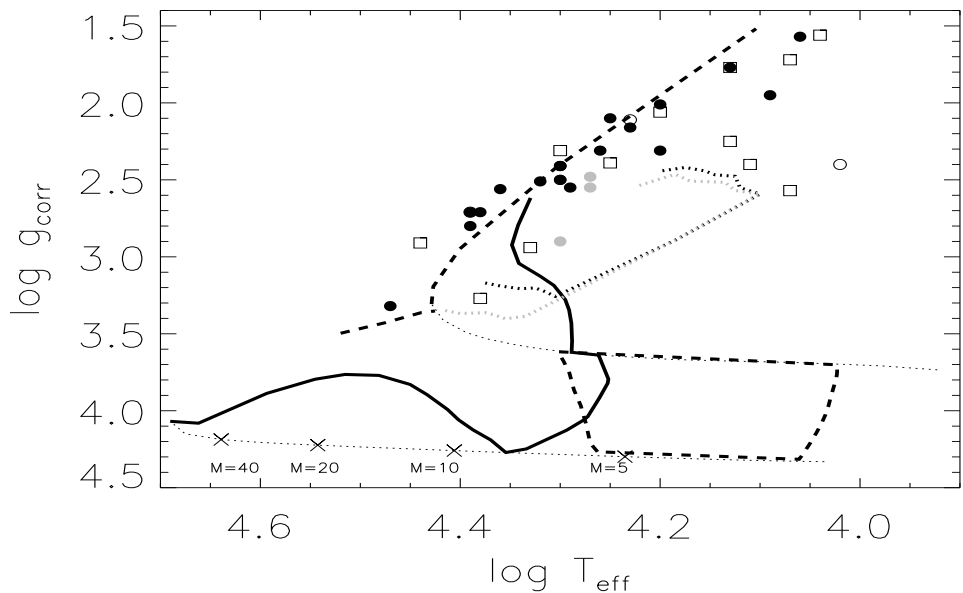}}} \end{center}
\caption[]{The position of B supergiants discovered to be periodically variable
from the Hipparcos mission is compared with instability computations for p~modes
(full lines) and g~modes (dashed lines) of main sequence models [12] and with
post-TAMS model predictions for $\ell=1$ (grey dotted) and $\ell=2$ (black
dotted) for B stars with masses up to 20\,M$_\odot$ computed by [54]. Figure
reproduced from Lefever et al.\ (2006) with permission from A\&A and from the
authors.}
\label{karolien}
\end{figure*}

Oscillations have not yet been firmly established in luminous stars with $\log
L/L_{\odot}>5$ and $M>20$\,M$_{\odot}$, although they are predicted in that part
of the HR diagram.  [12] and [54] predicted SPB-type g~modes to be unstable at
such high luminosities for respectively pre- and post-TAMS models
(Fig.\,\ref{karolien}).

[2] discovered a sample of B supergiants to be periodically variable with
SPB-type periods from the Hipparcos mission. These stars, and additional similar
ones, were subjected to detailed spectroscopic and frequency analyses by [55],
who found their masses to be below 40\,M$_\odot$ and photometric periods between
1 and 25\,d. The stars were found to be situated at the high-gravity limit of
$\kappa$-driven pre-TAMS g-mode instability strip ([12], see
Fig.\,\ref{karolien}).  This implies that the interpretation of their
variability in terms of nonradial g-mode oscillations excited by the
$\kappa\,$mechanism, as first suggested by [2], is plausible.

A new step ahead in the understanding of these stars was achieved by [54],
who detected both p and g~modes in the B2Ib/II star HD\,163899 from
MOST space-based photometry. The authors deduced 48 frequencies below
2.8\,d$^{-1}$ with amplitudes below 4\,mmag and computed post-TAMS stellar
models and their oscillation frequencies which turn out to be compatible with
the observed ones.  

Further research is needed to evaluate if seismic modelling in terms of internal
physics evaluation of these SPB supergiants, as [54] termed their target, is
feasible.

\section{Discussion and Future Prospects}

The classes of the $\beta\,$Cep and SPB stars are now well established,
containing more than 200 members each. Four of the brightest and slowest
rotators among the $\beta\,$Cep stars have been modelled seismically since 2003,
resulting in stringent constraints on the core overshoot parameter of $d_{\rm
ov}\in [0.05\pm 0.05,0.20\pm 0.05]$H$_{\rm p}$. Note that this range is lower
than the one found from a handful of eclipsing binaries with a B-type star
[56], implying that the latter probably also experience
rotational mixing near their core, which mimics additional core overshoot.  In
two stars (besides the Sun), seismic evidence for non-rigid internal rotation
was established. Both these stars have a core spinning faster than the envelope,
one with a factor three and the other one with a factor four. This was derived
from the computation of the Ledoux splitting coefficients, after successful
seismic modelling of the zonal components of observed frequency multiplets, and
a confrontation with the high-precision observed values of these coefficients.
We conclude that asteroseismology of $\beta\,$Cep stars has been highly
successful during the past few years, and its future looks very promising given
that several multisite campaigns of moderate rotators have been done but are not
yet exploited and CoRoT will be launched very soon.

Between one and five frequencies of g modes have been established in the
brightest among the SPB stars, from long-term photometric and spectroscopic
campaigns. This is rather disappointing, given the large observational effort
that went into this result. The example of the SPB star HD\,163830 observed by
MOST makes it clear that one needs photometry from space with a high duty cycle
to make efficient progress in the detection of frequencies for these stars. The
same holds true for the g modes in Be stars and B supergiants. We are eagerly
awaiting the results from CoRoT in this respect.

The oscillations detected in Be stars and very-late Oe stars show a multitude of
different behaviour, which is in full accordance with the one of $\beta\,$Cep
stars and SPB stars. It seems that pulsating Be stars are complicated analogues
of the SPB stars, while the $\zeta\,$Oph stars undergo the same oscillations
than $\beta\,$Cep stars, but the members of both these classes having emission
lines in their spectrum rotate typically above half of the critical velocity,
with some rotating very close to critical velocity. It remains to be studied
what the role of the oscillations is in the disk formation for the class of Be
stars as a whole.

Probing of B supergiant models has recently come within view, with the discovery
of nonradial g modes in such a star by the MOST mission. This case study is
complemented by the interpretation of the variability of the Hipparcos
lightcurves of a sample of some 40 B supergiants in terms of g modes. These two
entirely independent studies open the upper part of the HR diagram for seismic
tuning of stellar evolution models of supergiant stars, which are the precursors
of stellar black holes. At present, none of the existing analysis codes include
the effects of a radiation-driven stellar wind, which would be the next step
towards apropriate modelling of detected oscillation frequencies in such stars.

By far the largest stumbling block in the application of asteroseismology to
g-mode pulsators among the B stars is the lack of unambiguous mode
identification and good models including rotation in a consistent way. On the
observational side, this can only be resolved from coordinated initiatives,
because it requires long-term multisite multitechnique campaigns, including
multicolour photometry and high-resolution spectroscopy. Space photometry has
the potential of detecting a much higher number of oscillations than
ground-based photometry, as the MOST mission has shown us and will hopefully
continue to do so. However, it cannot deliver the badly needed mode
identification, because we do not have the comfort of dealing with frequency
spacings as in solar-like oscillators. Moreover, the rotational splitting is of
the same order or even larger than the separation between zonal g-mode
frequencies of subsequent radial order, implying that the measured frequency
spectrum is insufficient to unravel the nature of the detected modes.  On the
theoretical side, it is fair to state that we do not have appropriate seismic
models for stars rotating at a considerable fraction of their critical
velocity. Moreover, it was recently discovered that half of the SPB stars turn
out to have a magnetic field [57], such that not only the
Coriolis is important for such pulsators, but likely also the Lorentz force.

\section*{Acknowledgments}
The author is supported by the Fund for Scientific Research of Flanders (FWO)
under grant G.0332.06 and by the Research Council of the University of Leuven
under grant GOA/2003/04. She is very grateful to the organisers for giving her
the opportunity to present this work at the meeting.

\end{document}